\documentclass[fleqn]{article}
\usepackage{graphicx} % Required for inserting images
\usepackage[a4paper,top=25mm,bottom=25mm,left=25mm,right=25mm,nohead,nofoot,includeheadfoot]{geometry}
\usepackage{indentfirst}
\usepackage{multicol}
\usepackage{caption}
\usepackage{hyperref}
\usepackage[labelfont=bf]{caption}
\usepackage{amsmath}
\usepackage{physics}
\usepackage{float}
\usepackage{lipsum}
\usepackage[backend=biber, style=phys, sortcites=true, biblabel=brackets]{biblatex}
\begin{document}
\begin{center}
    {\LARGE\bfseries Hyperentanglement in Nanophotonic Systems with Discrete Rotational Symmetry \par}
    \vspace{1em}

    {\large\textbf{
    Lior Fridman\textsuperscript{1,2}, 
    Amit Kam\textsuperscript{2,3}, 
    Amir Sivan\textsuperscript{1,2}, 
    Guy Sayer\textsuperscript{1,2}, 
    Stav Lotan\textsuperscript{1,2}, 
    and Guy Bartal\textsuperscript{1,2}
    }\par}

    \vspace{1em}

    \begin{footnotesize}
    \textsuperscript{1}Andrew and Erna Viterbi Department of Electrical \& Computer Engineering, Technion, Haifa 32000, Israel\\
    \textsuperscript{2}Helen Diller Quantum Center, Technion – Israel Institute of Technology, Haifa 32000, Israel\\
    \textsuperscript{3}Physics Department, Technion – Israel Institute of Technology, Haifa 32000, Israel
    \end{footnotesize}
\end{center}

\vspace{0.5em}

\begin{abstract}

    We propose a scheme to generate hyperentanglement between photons carrying angular momentum in nanophotonic systems with discrete rotational symmetry. Coupling free-space photons into surface plasmon polaritons by a polygonal-shaped grating restricts the basis of the generated near-field modes to a finite set, thus creating a new mechanism for spatial mode entanglement. By encoding the incoming photons with spin and orbital angular momenta, we find that the system preserves the high-dimensional Hilbert space, in contrast to rotationally symmetric nanophotonic platforms, where the inseparability of spin and orbital degrees of freedom results in loss of information.
    We further show that by properly engineering the phase of the photons to conform to the polygonal boundary conditions, we achieve a new scheme for generating hyperentangled states, utilizing both the vector-field nature of the nanophotonic modes and the finite basis of states in polygonal boundary conditions.
    Our approach paves the way for on-chip quantum communication by expanding the Hilbert space used in computation.

\end{abstract}

\vspace{1em}

\begin{multicols}{2}
\section{Introduction}
Quantum nanophotonics leverages sub-wavelength control of light to enable on-chip quantum information processing\cite{madsen_quantum_2022,sibson_chip-based_2017}, single-photon sources\cite{silverstone_-chip_2014}, quantum sensing\cite{greenspon_designs_2025}, and networks\cite{ruskuc_multiplexed_2025}. The ability to engineer nanostructures to control how light couples into and emits out of a nano-photonic device provides a scalable platform for implementing photonic logic operations\cite{corrielli_rotated_2014,zhang_integrated_2018}. Such control also enables the exploitation of non-classical effects in compact circuits, laying the groundwork for integrated quantum technologies\cite{wang_integrated_2020}.

Entanglement lies at the core of those technologies, being a uniquely non-classical correlation between photons. Photons are exceptionally promising candidates for these tasks, as they can be entangled in multiple distinct degrees of freedom (DoF), including polarization\cite{kwiat_ultrabright_1999}, frequency\cite{olislager_frequency-bin_2010}, time-bin\cite{simon_creating_2005}, and angular momentum\cite{mair_entanglement_2001}.

Angular momentum constitutes a degree of freedom of particular interest; in free space, it appears as orbital angular momentum (OAM) and spin angular momentum (SAM)\cite{allen_iv_1999}. OAM, characterized by its helical phase front, forms an infinite-dimensional discrete Hilbert space\cite{molina-terriza_twisted_2007}, while SAM, tied to the light’s polarization, forms a two-level system\cite{beth_mechanical_1936}. However, tight confinement, as found in nanophotonic platforms, induces strong coupling between SAM and OAM\cite{bliokh_spinorbit_2015}, preventing their independent use for information encoding. 

Nanophotonic systems provide a platform to precisely control and harness this strong coupling between the spin and orbital angular momenta, offering two key advantages for engineering complex photonic states. The first is the ability to carefully engineer the geometry of the coupling from free-space to the nanophotonic platform. This geometry applies specific boundary conditions that determine the spatial profile of the nanophotonic states.\cite{gorodetski_observation_2009,tsesses_spinorbit_2019,tsesses_optical_2018}. Second, the tight focusing inherent to these devices breaks the paraxial approximation, thus making nanophotonic states vector fields. This property was recently exploited to entangle photons in their total angular momentum (TAM) within a nanophotonic system\cite{kam_near-field_2025} and to generate angular momentum-based qudits\cite{kam_tracking_2025}. 

Leveraging the two key advantages of nanophotonic systems in tandem provides new opportunities for hyperentanglement\cite{kwiat_hyper-entangled_1997} -  simultaneous entanglement of photons in multiple DoFs by initially unentangled photons\cite{nemirovsky-levy_increasing_2024, suo_generation_2015, barreiro_generation_2005}, within a compact nanophotonic device for scalable on-chip high-dimensional quantum protocols\cite{nemirovsky-levy_expanding_2024}.

Here, we exploit the coupling between free-space photons and a nanophotonic platform to increase the Hilbert space and introduce new schemes for on-chip hyperentanglement. We show that breaking the rotational symmetry of nanophotonic modes preserves the quantum information that is typically lost during the coupling to fully symmetric nanophotonic states\cite{kam_near-field_2025}. We further propose and analyze a complete scheme to generate hyperentanglement by engineering the incoming photons to match the discrete rotational symmetry of the polygonal boundary conditions.  

Generating complex quantum states in nanophotonic platforms offers a route for engineering on-chip sources of multipartite and high-dimensional entanglement, with implications for future theoretical and experimental works.

\section{Surface states}
In nanophotonics systems, the SAM and OAM are known to merge such that neither can be observed separately, and only the TAM can be measured. Namely, photons imprinted with field rotation $\sigma$ (polarization handedness)  and phase rotation $e^{jl\theta}$  ($l$ quantifies the orbital angular momentum - the number of times an azimuthal phase completes a cycle of $2\pi$) become definable only by the TAM $n = l+\sigma$ where the separate contribution of the imprinted spin and OAM can no longer be distinguished.

We consider a slab-based nanophotonic system, e.g., surface plasmons on a metal-air interface, carved with a polygonal slit or grating coupler as introduced in \cite{tsesses_spinorbit_2019}. In such systems, circularly-polarized light impinging on the coupling grating results in a vector near-field whose out-of-plane field component can be regarded as an interference of N plane waves -

\begin{equation}\label{eq:Field expansion}
    E_z=e^{-|k_z|z}\sum_{n=1}^{N}A_{n}e^{j\phi_n}e^{-jk_{sp}[cos(\theta_n)x+sin(\theta_n)y]}
\end{equation}

Here, x,y are the in-plane coordinates measured with respect to the center of the slit; each side lies at an angle \(\theta_n = \frac{2\pi}{N}n\) for \(n = 1,2,...,N\) where N is the total number of sides. The coefficients $A_n$ and $\phi_n$ represent the amplitude and relative phase delay of each of the plane waves, respectively \cite{tsesses_optical_2018,tsesses_spinorbit_2019}.

Such N-sides polygon excitation slit supports a finite eigen-mode basis of dimension N with discrete rotational symmetry (see supplementary). As such, these modes can be characterized by a topological number representing the number of times the phase completes a cycle of $2\pi$ about the center of the unit cell, similarly to the topological charge in circularly symmetric systems. For phase arrangement satisfying $\phi_n=\theta_n\cdot l$, this number is an integer and is limited to \cite{tsesses_spinorbit_2019} -

{\small   
\begin{equation}
\begin{aligned}
n ={}&\,
\begin{cases}
(l + \sigma)\bmod N,     & \text{if }(l + \sigma)\bmod N \le \tfrac{N}{2},\\[4pt]
(l + \sigma)\bmod N - N, & \text{if }(l + \sigma)\bmod N > \tfrac{N}{2}
\end{cases}
\end{aligned}
\label{eq:mode-relation}
\end{equation}
}

The in-plane electric field in such geometry is derived using Maxwell's equations and, in the rotating field basis, is in the form of - 

\begin{equation}\label{eq: In-plane fields}
    \binom{E_{\sigma_-}}{E_{\sigma_+}}\propto \sum_{n=1}^{N} \binom{e^{j (\phi_{n}-\theta_n)}}{e^{j (\phi_{n}+\theta_n)}} e^{-j k_{sp}\left[\cos \left(\theta_{n}\right) x+\sin \left(\theta_{n}\right) y\right]}
\end{equation}

The left- and right-handed rotating field components can also be characterized by the topological number corresponding to the number of phase rotations about the center,  which differs in \(\pm1\), respectively,  from that of the out-of-plane component. These classical relations, obtained directly from Maxwell's equations, remain applicable in the single-photon regime.

Introducing the notation $\ket{n}$ for the mode with TAM (equal to the above-mentioned topological number) of $n$, we derive the following relation between the free-space angular momentum quantum numbers to the near-field number -

\begin{multline}
    \ket{l}\ket{\sigma}
    \overset{\text{Near Field}}{\xrightarrow{}}\ket{n} \\
    \overset{\text{Field Components}}{\xrightarrow{}} 
    \ket{\sigma_{+}}\ket{n-1}-\ket{\sigma_-}\ket{n+1}
    \label{eq:AM transformation}
\end{multline}
Namely, a photon carrying a spin $\ket{\sigma}$ and OAM  $\ket{l}$ is transformed to a vector near-field state of $\ket{n}$ with in-plane field components of $\ket{\sigma_{+}}\ket{n-1}-\ket{\sigma_-}\ket{n+1}$. We note that the out-of-plane component has no SAM; therefore, its OAM is always equal to the TAM.

\section{Hyper-entangling scheme}
We first investigate Hyper-entanglement in nanophotonic systems excited by a plane wave carrying no OAM ( $l=0$). In this case, the angular momentum is contributed solely by the SAM, yet its contribution cannot be distinguished from that of the OAM \cite{kam_near-field_2025}. While this indistinguishability appears to lose information, it opens up an alternative channel for entanglement \cite{kam_tracking_2025}. We aim at leveraging this channel, along with the newly-found attributes of discrete rotational symmetry in nanophotonics, to introduce a new scheme for hyperentanglement.

Hyperentangled photonic states are generally produced via a two-photon state, e.g., $\ket{H}\ket{V}$,  generated in a spontaneous parametric down-conversion (SPDC) process. Following \autoref{eq:AM transformation}, the quantum state of such nanophotonic platform can be described as a N00N state in the TAM

{\makeatletter\@fleqnfalse
\begin{equation}
    \ket{HV}\xrightarrow{\text{Near Field}}\ket{+1}^{\otimes2}-\ket{-1}^{\otimes2}
\end{equation}
\makeatother}

 where $\ket{n=\pm 1}$ is the TAM of each state. Taking into account the in-plane field components \autoref{eq:AM transformation}, the N00N state is decomposed into - 

\begin{equation}
\begin{aligned}
   \ket{+1}^{\otimes 2} - \ket{-1}^{\otimes 2} 
   &\stackrel{\text{Field Components}}{\longrightarrow} \\
   &[\ket{\sigma_+}\ket{0} - \ket{\sigma_-}\ket{+2}]^{\otimes 2} \\
   \phantom{\stackrel{\text{Field Components}}{\longrightarrow}}
   -&[\ket{\sigma_+}\ket{-2} - \ket{\sigma_-}\ket{0}]^{\otimes 2}
\end{aligned}
\label{eq:hyper-state}
\end{equation}

This relationship is especially significant in discrete rotational symmetry where the mode basis becomes cyclic; specifically, for a square coupling slit with N=4, only 4 eigenmodes are supported \cite{spektor_spin-patterned_2015} where the cyclic nature of the basis (\autoref{eq:mode-relation}), dictates that $\ket{+2} \equiv \ket{-2}$. This aliasing-like behavior, in which higher-order modes map onto lower-order ones, introduces an additional entanglement mechanism.

\end{multicols}

\begin{figure}[H]
    \centering
    \includegraphics[width=\linewidth]{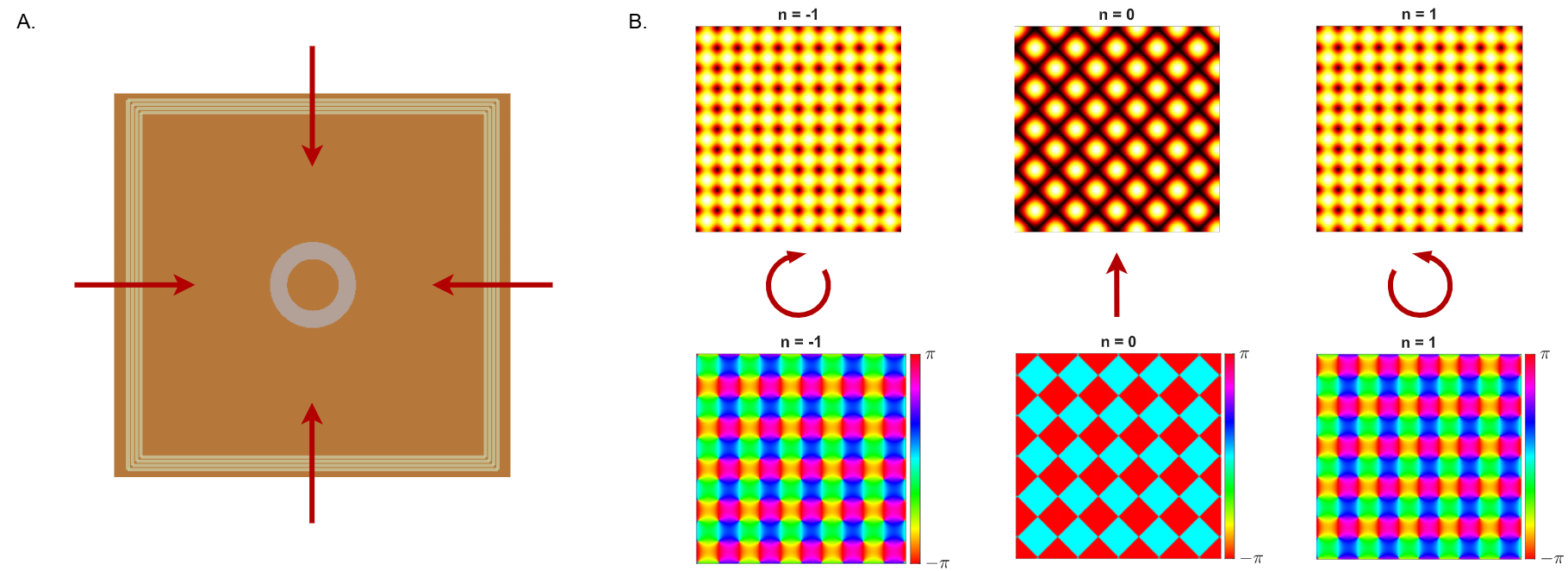}
    \caption{\textbf{Eigenmodes of square boundary conditions} - (a) An excitation from a square slit is modeled as 4 plane waves propagating along the slab nanophotonic system. (b) The field distribution of each field component.  top - amplitude,  bottom - phase.}
    \label{fig: square modes}
\end{figure}

\begin{multicols}{2}

Upon scattering to free-space \cite{kam_near-field_2025}, this entanglement takes the form of -

\begin{equation}
    \begin{aligned}
        \ket{\psi}
          &= \bigl[\ket{\sigma_+}^{\otimes^2} - \ket{\sigma_-}^{\otimes^2}\bigr]
             \otimes
             \bigl[\ket{A}^{\otimes^2} - \ket{B}^{\otimes^2}\bigr] \\[6pt]
          &-\ket{\sigma_+}\ket{A}\ket{\sigma_-}\ket{B}-\ket{\sigma_-}\ket{B}\ket{\sigma_+}\ket{A} \\
          &+\ket{\sigma_+}\ket{B}\ket{\sigma_-}\ket{A}+\ket{\sigma_-}\ket{A}\ket{\sigma_+}\ket{B}
    \end{aligned}
\label{eq:final state}
\end{equation}

\begin{figure}
    \centering
    \includegraphics[width=0.55\linewidth]{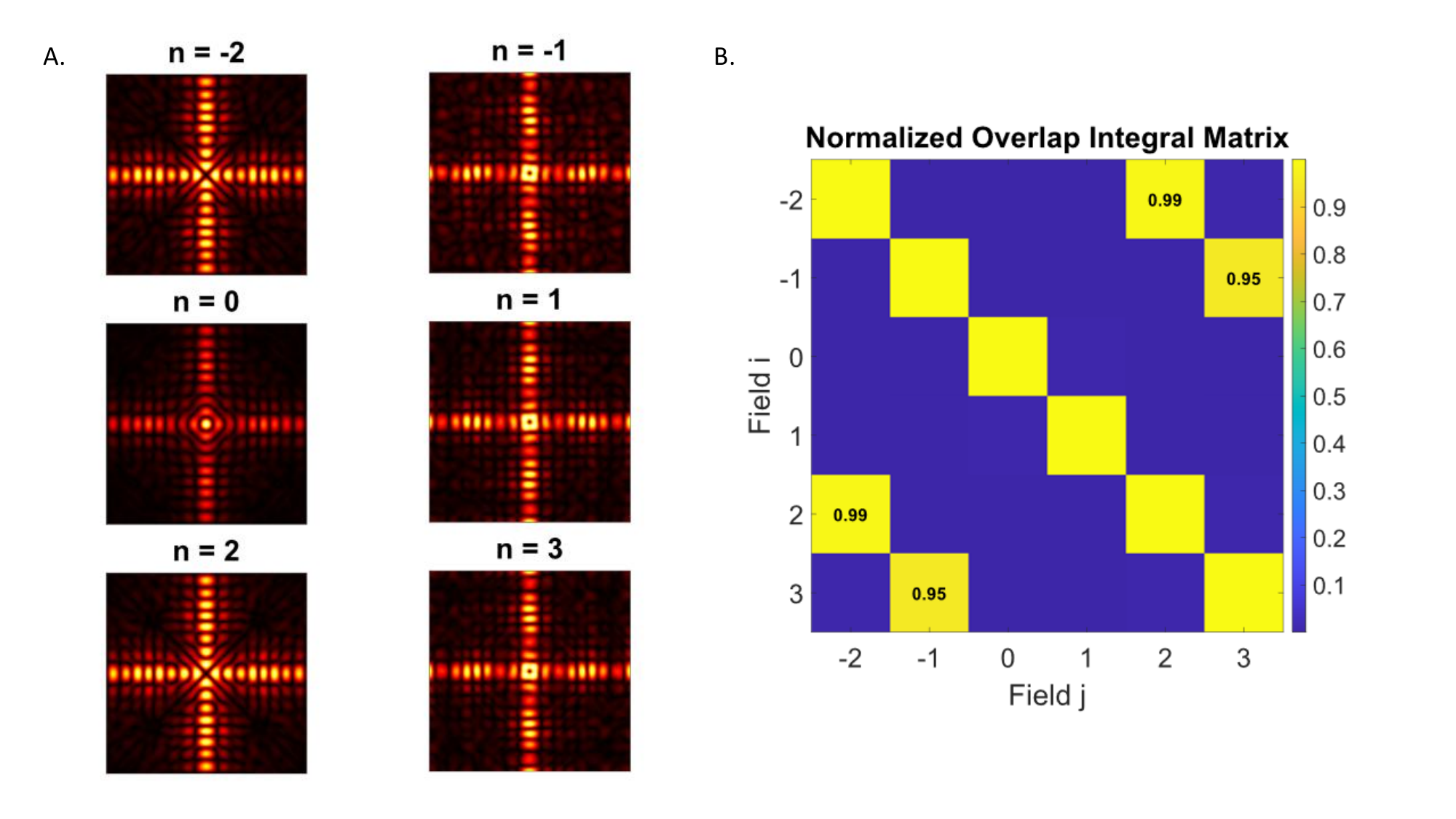}
    \caption{\textbf{Out-coupled modes} - a.  the far-field (Fourier) plane amplitude of different out-coupled modes. b.  calculated overlap integrals between the different modes.
  .}
    \label{fig:out-coupeled fields}
\end{figure}

Where $\ket{A}=\ket{0}$ and $\ket{B}=\ket{2}\equiv\ket{-2}$ are the spatial profiles of the out-coupled field, and are obtained by scattering out the $\ket{0}$ and $\ket{-2}$ nanophotonic modes \cite{kam_tracking_2025}. This scattered state is hyperentangled in SAM and a spatial DoF, which is directly derived from the nanophotonic eigenmodes \autoref{fig:out-coupeled fields}. 

We employ a finite-difference time-domain (FDTD) numerical simulation to model the in- and out-coupling processes. Such simulations are based on Maxwell's equations and are classical in essence, yet can well simulate the dynamics of a single photon in our system. We specifically aim at showing that the outcoupling scheme maintains orthogonality, thereby allowing for further free-space quantum computation using the resultant state.

The simulation accounts for a \(220nm\) layer of gold on top of a BK7 substrate, where the excitation slit (in-coupler) is carved  through the gold. The out-coupler is assumed to be circular, carved at a depth of \(80nm\). The phase structure of the exciting wave is of the form  $\phi_n=\theta_n\cdot l$, where $l$ is an integer that we vary to cover all possible eigenmodes (-2,-1,0,1,2,3). We then calculate the overlap between the different modes, "measured" in the far-field after being scattered by a circular slit.

\autoref{fig:out-coupeled fields} depicts the far-field (Fourier plane) amplitude of the out-coupled field for different $l$ values.  While the spatial form of  the far-field is not identical to the nanophotonic modes, it retains the phase structure in the center for the pattern, which corresponds to the topological  charge, such that, for each $l$ value, the out-coupled photon can still be described by \(\ket{\psi} = \ket{\sigma_+}\ket{l}-\ket{\sigma_-}\ket{l+2}\). Moreover, calculating the overlap integrals between the fields scattered for different values of $l$, reveals that the orthogonality between the different modes is also preserved. 

Importantly, the cyclical nature of the basis is maintained for the scattered modes. As a result, analogous to the near-field case, the scattered state exhibits hyperentanglement. Thus, the simulation results suggest that a hyperentangled state is created by a free-space photon pair at their far-field.

\end{multicols}

\begin{multicols}{2}

This scheme can be generalized for use with various polygonal slits by tailoring the phase front of the initial excitation to comply with the slit's discrete rotational symmetry.  This relation is expressed as $\phi_n=\theta_n\cdot l$ where  $l$ represents the topological charge of the excitation itself, which can be a single integer value or an equal superposition of multiple charges.

Applying such a tailored excitation leads to a general condition that must be met to achieve hyperentanglement: the resulting in-plane field distribution must satisfy the equivalence -

{\makeatletter\@fleqnfalse
\begin{equation}
    \ket{l+2} \equiv \ket{l-2} 
    \label{eq:hyper_req}
\end{equation}
\makeatother}

This requirement places a constraint on the number of sides $N$ of the polygonal slit. Specifically, it can be fulfilled only when $N=4m$ (with m an integer), in which case the excitation must involve an equal superposition of m topological charges. This condition stems from the intrinsic splitting of the topological charge across the different field components (\autoref{eq:AM transformation}). 

For a simple case, such as a square slit (N=4), this condition is naturally satisfied. For other polygons, however, a more complex, structured phase front is necessary, requiring the excitation to be in a superposition of topological charges.

For instance,  an octagonal slit (N=8) requires  a superposition state of $\ket{l}=\ket{-1}+\ket{3}$. Substituting this state in \autoref{eq:hyper_req} yields -
\begin{align*}
    &\ket{l+2}=\ket{-1+2}+\ket{3+2}=\ket{1}+\ket{5} \\
    &\ket{l-2} = \ket{-1-2}+ \ket{3-2} = \ket{-3}+\ket{1}
\end{align*}
According to \autoref{eq:mode-relation} , we find that $\ket{5}\equiv\ket{-3}$ the condition is thus satisfied.

\begin{figure}[H]
    \centering
    \includegraphics[width=0.5\linewidth]{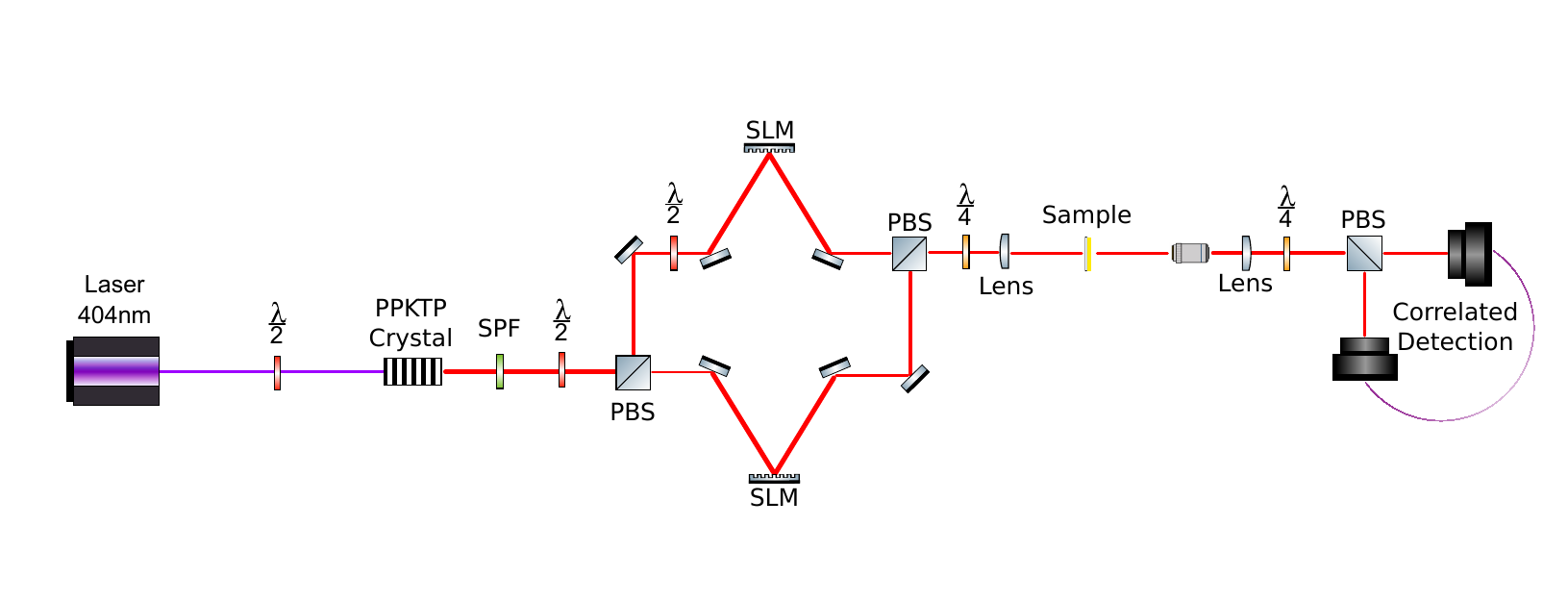}
    \caption{\textbf{Experimental system suggestion} - $404_{nm}$ laser source is down-converted into a collinear photon pair. We split the photons based on polarization into a dual SLM configuration, each imprinting an identical spatial profile according to the sample excitation slit. Both photons are then combined and coupled to the nanophotonic system. Using another polarizing beam splitter, we then split the generated state and perform a correlation measurement in the selected basis.}
    \label{fig: Exp system}
\end{figure}

The resultant hyperentangled state can be experimentally measured through post-selection, using parity measurements of SAM and the spatial DoF facilitated by single-photon detectors.  \autoref{fig: Exp system}, shows a  proposed  proof-of-concept  experimental scheme for generating and measuring the hyperentangled state. 

In this arrangement, we propose a metal-dielectric interface as our host nanophotonic platform whose eigenmodes are surface plasmon polaritons (SPPs). The principles of exciting and modeling these modes are well-established \cite{barnes_surface_2003, raether_surface_1988,zayats_nano-optics_2005}. In our scheme, the SAM entanglement is converted into distinct spatial paths using a polarizing beam splitter (PBS); alternatively, owing to the state’s symmetry, a match-filter could be used to map the spatial DoF into the path DoF.  To isolate the hyperentangled state, the scheme relies on post-selection, achieving a success probability of 0.5, as shown in \autoref{eq:final state} via correlated measurements.

\end{multicols}

\begin{multicols}{2}

\section{Isomorphic mapping}
We next propose a scheme to retain the indistinguishability between the photon angular momentum DoFs that is typically lost upon coupling to nanophotonic structures. 
Consider an incident photon carrying both SAM and OAM, e.g., a Laguerre-Gauss mode, with a phase front - $$\phi_n= l \theta$$
Here $l$ is the topological number that quantifies the OAM of the excitation.

The SAM and OAM together define the mode generated in the nanophotonic system, as was shown in  \autoref{eq:mode-relation}. According to \autoref{eq:Field expansion}, the quantum number that defines the mode is determined by the relative phases of the incident wave at each side of the polygon, which contains contributions from both the SAM and OAM.

The SAM contribution stems from the temporal rotation of the E-field and the dipolar nature of the excitation, which only couples the field component  perpendicular to the slit \cite{barnes_surface_2003}. Importantly, each side is excited uniformly across its length, meaning that the SAM does not introduce any relative temporal phase delay along a single side. 

Conversely, the OAM imposes both a phase between the sides of the polygonal coupler and a phase gradient along the length of each side. While the topological charge $l$, together with the polarization handedness, determines the  TAM of the excited field, which serves as the quantum number of the state, the phase gradient across each side rotates the entire mode about the coupler center. Consequently, the excited mode is characterized not only by its TAM but also by this static rotation angle, enabling the distinction between different angular momentum contributions. 

To estimate the static rotation angle of the mode, we retain the first-order term in the Taylor expansion of the phase delay along each side, modeling it as an effective prism that imparts a linear phase delay. Under this approximation, the static rotation angle can be expressed as -

{\makeatletter\@fleqnfalse
\begin{equation}
\label{eq:approx-rotation}
\theta_{\mathrm{stat}}{\left(l\right)}
= \arctan\!\left(\frac{2\,l}{k_{sp}\,L}\right)
\end{equation}
\makeatother}

This expression provides a one-to-one mapping between the OAM charge $l$ and the mode rotation for a given slit (specified by $L$, which is derived from the number of sides $N$), providing a simple analytical estimation of the static rotation angle.

\autoref{eq:approx-rotation}  provides two useful limitations:

\begin{itemize}
  \item For $l=0$ (no OAM), the static angle vanishes -
  $$
  \theta_{\mathrm{stat}}(l=0)=0.
  $$
  \item In the limit $L\to0$ (the polygonal slit approaches a continuous circle as the number of sides $N\to\infty$), the argument of the arctangent diverges for fixed nonzero $l$, hence -
  $$
  \lim_{L\to 0}\theta_{\mathrm{stat}}=\mathrm{sign}(l)\cdot\frac{\pi}{2},
  $$
\end{itemize}

\begin{figure}[H]
    \centering
    \includegraphics[trim = 0pt 0pt 0pt 75pt ,width=0.7\linewidth]{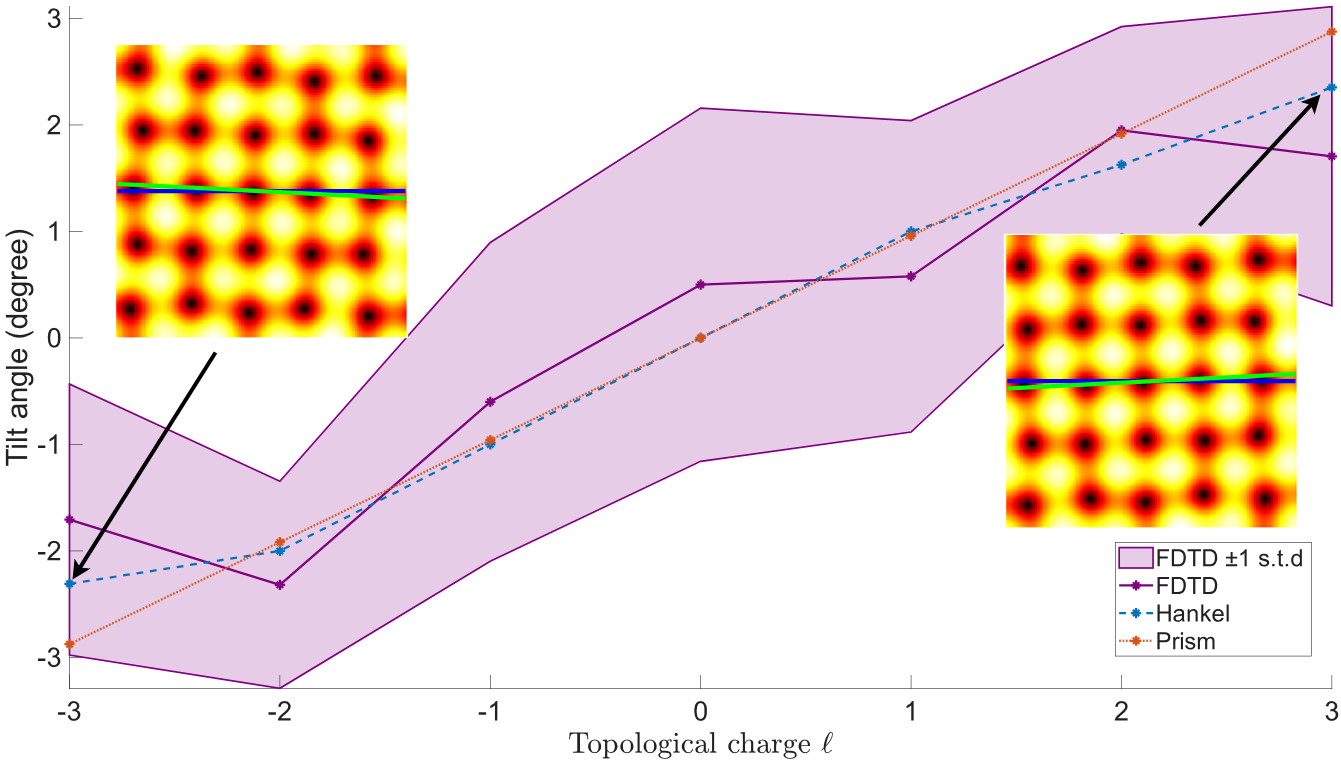}
    \caption{\textbf{Numerical tilt angle calculation} - Tilt angle of the excited modes as a function of the excitation topological charge for a square slit with side length $L=15_{\mu m}$. The lines represent tilt angles obtained by different methods: orange — prism approximation, blue — Huygens principle model, and purple — FDTD simulation, with the shaded band indicating ±1 circular standard deviation. Insets depict examples of rotated in-plane fields, with a blue horizontal line shown for reference and a green line indicating the tilt angle. }
    \label{fig: Rotations}
\end{figure}

\autoref{fig: Rotations},  presents the static rotation angle dependence on the topological charge for N=4, obtained by three different approaches: the effective prism approximation, Huygens principle calculation \cite{soifer_diffractive_2016},  and a full FDTD simulation. The prism approximation shows excellent agreement with the more rigorous Huygens principle, while the FDTD deviation results both from the numerical angle-extraction using the Hough transform and from intrinsic modeling differences between the discretized FDTD simulation and the analytical models.

\end{multicols}

\begin{multicols}{2} 

We now define the static rotation operator based on the static rotation angle as two translation operators with respect to the in-plane axis - 
{\makeatletter\@fleqnfalse
\begin{equation}
    \label{eq:Rotation_op}
    \begin{split}
    & R\left( z,\theta_{stat} \right)=T_x(-y\cdot \theta_{stat})T_y(x\cdot\theta_{stat})
    \end{split}
\end{equation}
\makeatother}
The operator eigenmodes are the in-plane fields rotated by $\theta_{stat}$. When both rotational contributions are included, the excited states can be written as the product  $\ket{n}\ket{\theta_{stat}}$.  In systems possessing discrete rotational symmetry, a measurement of the static angle reveals the excitation’s OAM \autoref{eq:Rotation_op}, while a measurement of the field profile determines the TAM \autoref{eq:mode-relation}, establishing an isomorphism between free-space angular momentum and the near-field excitation.  Consequently, measuring any single plasmonic field component is sufficient to determine the excitation’s SAM and OAM independently.  

\section{Conclusions}
To conclude, we proposed a new platform for hyperentanglement generation, based on the discrete rotational symmetry of  polygonal boundary conditions, the vectorial nature of nanophotonic states, and the angular momentum of the incoming photon.  In addition, we presented a nanophotonic scheme that preserves the DoFs of free space beyond the paraxial approximation and derived an analytical approximation for the static rotation angle of the in-plane field in the case of OAM-carrying excitations, establishing an isomorphic mapping. These results pave the way toward scalable, on-chip quantum photonic platforms by providing a robust mechanism for generating complex entangled states which are necessary for efficient quantum computation. 

\section*{Data Availability}

The data used in this study is available from the corresponding author upon reasonable request.

\section*{Code Availability}

The code used in this study is available from the corresponding author upon reasonable request.

\section*{Acknowledgements}
The authors thank Meir Orenstein for fruitful discussions. 
This research was supported by the Israel Science Foundation (ISF), Grant No. 3620/24 and the Israel Ministry of Innovation, Science and Technology, Grant No. 2033419. A.K. acknowledges the support from the Azrieli fellowship and the support from the Helen Diller Quantum Center at the Technion. A.S acknowledges the support from the Helen Diller Quantum Center at the Technion. G.S. acknowledges the support from the Russel Berrie Nanotechnology Institute (RBNI) at the Technion. 

\section*{Author Contributions}

L.F. conceived the project. L.F. and A.S. performed the FDTD simulation, L.F., G.S., and S.L. analyzed the data, L.F. and A.K. derived the theoretical model. G.B. supervised the research. All authors discussed the results and participated in writing the manuscript.

\section*{Competing Interests}
The authors declare no competing interests.

\printbibliography
\end{multicols}
\end{document}